\def\del{\partial}
\def\half{{1\over2}}

\def\dslash{\del\kern-0.55em\raise 0.14ex\hbox{/}}
\def\abs#1{{\left|{#1}\right|}}
\def\expecv#1{\langle #1 \rangle}
\def\tpsi{{\tilde\psi}}

\newcommand{\PRD}[3]{Phys. Rev. {\bf D{#1}} (19{#2}) {#3}}

\newcommand{\NPB}[3]{Nucl. Phys. {\bf B{#1}} (19{#2}) {#3}}

\newcommand{\PTP}[3]{Prog. Theor. Phys. {\bf {#1}} (19{#2}) {#3}}

\documentstyle[12pt,epsf]{article}
\makeatletter

\@addtoreset{equation}{section}
\makeatother
\begin{document}
\begin{titlepage}
\begin{flushright}
SAGA--HE--99\\
February 10,~1996
\end{flushright}
\vspace{50pt}
\centerline{\Large{\bf Numerical Approach to CP-Violating Dirac 
Equation}}
\vskip2.0cm
\begin{center}
{\bf Koichi~Funakubo$^{a,}$\footnote{e-mail: funakubo@cc.saga-u.ac.jp},
 Akira~Kakuto$^{b,}$\footnote{e-mail: kakuto@fuk.kindai.ac.jp},\\
 Shoichiro~Otsuki$^{b,}$\footnote{e-mail: otks1scp@mbox.nc.kyushu-u.ac.jp}
 and Fumihiko~Toyoda$^{b,}$\footnote{e-mail: ftoyoda@fuk.kindai.ac.jp}}
\end{center}
\vskip 1.0 cm
\begin{center}
{\it $^{a)}$Department of Physics, Saga University,
Saga 840 Japan}
\vskip 0.2 cm
{\it $^{b)}$Department of Liberal Arts, Kinki University in Kyushu,
Iizuka 820 Japan}
\end{center}
\vskip 1.5 cm
\centerline{\bf Abstract}
\vskip 0.2 cm
\baselineskip=15pt
We propose a new method to evaluate the chiral charge flux, which is
converted into baryon number in the framework of the charge transport
scenario of electroweak baryogenesis.
By the new method, one can calculate the flux in the background of
any type of bubble wall with any desired accuracy.
\end{titlepage}
\baselineskip=18pt
\setcounter{page}{2}
\setcounter{footnote}{0}
\section{Introduction}
In the scenario of electroweak baryogenesis\cite{reviewEB}, 
the chiral charge flux, which is converted to baryon number by
the sphaleron process, is one of key ingredients.
That is supposed to be generated by two mechanisms, both of which
depend on the interaction of the fermions with the CP-violating background
of the bubble wall created at the electroweak phase transition (EWPT). 
One of the mechanisms is the charge transport scenario, which is 
quantum mechanical and effective for thin walls, the other is 
the spontaneous baryogenesis, which is classical and effective for 
thick walls.\par
As for the charge transport scenario, the flux is determined by the
difference in the reflection coefficients of the left- and right-handed
fermions, as well as by that in the distribution functions in the broken
and the symmetric phases divided by the bubble wall.
The reflection coefficients are calculated by solving the Dirac 
equation in the background of the bubble wall, which is accompanied
spatially varying CP violation.
Cohen, {\it et al.}\cite{NKC} first evaluated the flux by solving 
the Dirac equation numerically, assuming the kink-type profile for 
the modulus of the Higgs field and a function linear in the kink for
the phase. On the other hand, we formulated a perturbative method
applicable for small CP-violating phase and proved various relations
among the reflection and transmission coefficients\cite{FKOTTa,FKOTTb}.
Although it enables us to derive analytical relations, it can actually
be applied only to the case in which the unperturbed Dirac equation is
exactly solved and the CP-violating phase is sufficiently small.
However the profile of the bubble wall should be determined by the
dynamics of the EWPT.
We pointed out, by solving the equations of motion for the gauge-Higgs
system, that the CP-violating phase can be of $O(1)$ even if that is
small in the broken phase at $T=0$, according to the parameters in the 
effective potential\cite{FKOTTc}.
For a larger CP-violating phase, we naively expect larger baryon number
to be generated.
So we need a method to evaluate the chiral charge flux for arbitrary
profile of the bubble wall.\par
In \cite{NKC}, the one-dimensional Dirac equation is solved within
a finite range $[0,z_0]$ including the bubble wall, 
imposing the boundary condition in which plane waves are injected 
at the boundary.
Strictly, a plane wave is not an eigenfunction at any finite $z$,
so that the choice of the range will affect the precision of the
results.
In this paper, we propose another procedure to solve the Dirac 
equation and to calculate the reflection and transmission 
coefficients. Although the method is numerical, its precision can
be so controlled that one can obtain the results with any accuracy.
In section 2, we write down the CP-violating Dirac equation and
give the expressions of the reflection and transmission coefficients
of the chiral fermions. The procedure of the numerical analysis
is given in section 3, and it is applied to some profiles, including
those which we found in \cite{FKOTTc} and that in \cite{NKC} for 
comparison, in section 4. The final section is devoted to summary and
discussions.
\section{CP-Violating Dirac Equation}\noindent
\subsection{The Dirac equation}
We consider a fermion in the background of a complex scalar, to which
it couples by a Yukawa coupling. For the scalar to have a nontrivial
complex vacuum expectation value (VEV), we assume it is a part of 
an extended Higgs sector of the electroweak theory, such as MSSM or
more general two-Higgs-doublet model.
At the electroweak phase transition, the Higgs changes its VEV from 
zero to nonzero value near the bubble wall created if the phase 
transition is first order.\par
Then the Dirac equation describing such a fermion is
\begin{equation}
 i\dslash\psi(x)-m(x)P_R\psi(x)-m^*(x)P_L\psi(x) = 0,
     \label{eq:Dirac1}
\end{equation}
where $P_R={{1+\gamma_5}\over2}$,$P_L={{1-\gamma_5}\over2}$ and 
$m(x)= -f\expecv{\phi(x)}$ is a complex-valued function of spacetime.
Here $f$ is the Yukawa coupling.
When the radius of the bubble is macroscopic and the bubble is
static or moving with a constant velocity, we can regard $m(x)$ as
a static function of only one spatial coordinate:
$$
   m(x) = m(t,{\bf x}) = m(z).
$$
Putting
\begin{equation}
 \psi(x) = {\rm e}^{i\sigma(-Et+{\bf p}_T\cdot{\bf x}_T)},\qquad
 (\sigma=+1\mbox{ or } -1)
  \label{eq:separation}
\end{equation}
(\ref{eq:Dirac1}) is reduced to
\begin{equation}
 \left[ \sigma(\gamma^0E-\gamma_Tp_T)+i\gamma^3\del_z
    -m_R(z) - i\gamma_5 m_I(z) \right] \psi_E({\bf p}_T,z) = 0,
   \label{eq:Dirac2}
\end{equation}
where
\begin{eqnarray*}
 {\bf p}_T&=&(p^1,p^2),\qquad {\bf x}_T=(x^1,x^2),
 \qquad p_T = \abs{{\bf p}_T},\qquad
 \gamma_T p_T = \gamma^1 p^1 + \gamma^2 p^2,  \\
 m(z) &=& m_R(z) + i m_I(z).
\end{eqnarray*}
If we denote $E=E^*\cosh\eta$ and $p_T=E^*\sinh\eta$ with
$E^*=\sqrt{E^2-p_T^2}$, ${\bf p}_T$ in (\ref{eq:Dirac2}) can be
eliminated by the Lorentz transformation
$\psi\mapsto\psi^\prime = {\rm e}^{-\eta\gamma^0\gamma_5}\psi$:
\begin{equation}
 \del_z\psi_E(z)=i\gamma^3\left[-\sigma E^*\gamma^0
                 +m_R(z) + i\gamma_5 m_I(z) \right]\psi_E(z).
	\label{eq:Dirac3}
\end{equation}
In our work on the perturbative treatment of CP violation, we
used the eigenspinor of $\gamma^3$, since unperturbative part of
the Dirac equation contains only $\gamma^3$.
Here we shall use the chiral representation of the $\gamma$-matrices
as in \cite{NKC}. So the expression of the Dirac equation is the same
as that in \cite{NKC}, but we write it for self-containedness.\par
In the chiral basis, the $\gamma$-matrices are represented as
$$
 \gamma^0=\pmatrix{0&-1\cr -1&0\cr},\quad
 \gamma^k=\pmatrix{0&\sigma_k\cr-\sigma_k&0\cr},\quad
 \gamma_5=\pmatrix{1&0\cr0&-1\cr}.
$$
If we write the four-spinor as
$$
 \psi_E(z) = \pmatrix{\psi_1(z)\cr \psi_2(z)\cr \psi_3(z)\cr \psi_4(z)\cr},
$$
the Dirac equation is decomposed into two two-component equations:
\begin{eqnarray}
 \del_z\pmatrix{\psi_1(z)\cr \psi_3(z)\cr} &=&
 i\pmatrix{\sigma E^* & m^*(z) \cr -m(z) & -\sigma E^*\cr}
  \pmatrix{\psi_1(z)\cr \psi_3(z)\cr},   \label{eq:Dirac4-1}  \\
 \del_z\pmatrix{\psi_4(z)\cr \psi_2(z)\cr} &=&
 i\pmatrix{\sigma E^* & m(z) \cr -m^*(z) & -\sigma E^*\cr}
  \pmatrix{\psi_4(z)\cr \psi_2(z)\cr}.   \label{eq:Dirac4-2}
\end{eqnarray}
\par%
Now let us introduce dimensionless parameters.
Suppose that $a$ is the parameter of mass dimension, whose inverse
characterizes the width of the bubble wall. 
We shall comment on practical choice of it in the later section.
Then put
\begin{eqnarray}
 x&\equiv& az,    \nonumber \\
 \epsilon&\equiv& E^*/a,            \label{eq:dimensionless1}  \\
 \xi(x){\rm e}^{i\theta(x)}&\equiv& m(z)/a \nonumber
\end{eqnarray}
Since we are interested in the bubble wall background, the
$m(z)$ is supposed to behave as
\begin{eqnarray*}
  m(z) &\rightarrow& m_0{\rm e}^{i\theta_0},\qquad\mbox{as }
  z\rightarrow +\infty, \\
  m(z) &\rightarrow& 0,\qquad\mbox{as } z\rightarrow -\infty,
\end{eqnarray*}
where $m_0$ is the fermion mass and $\theta_0$ is the CP-violating
angle in the broken phase at the phase transition.
We also put
\begin{equation}
  \xi \equiv m_0 / a.               \label{eq:dimensionless2}
\end{equation}
In terms of these dimensionless parameters, the Dirac equation
is expressed as
\begin{eqnarray}
 \del_x\pmatrix{\psi_1(x)\cr \psi_3(x)\cr} &=&
 i\pmatrix{\sigma\epsilon & \xi(x){\rm e}^{-i\theta(x)} \cr
           -\xi(x){\rm e}^{i\theta(x)} & -\sigma\epsilon \cr}
  \pmatrix{\psi_1(x)\cr \psi_3(x)\cr},   \label{eq:Dirac5-1}  \\
 \del_x\pmatrix{\psi_4(x)\cr \psi_2(x)\cr} &=&
 i\pmatrix{\sigma\epsilon & \xi(x){\rm e}^{i\theta(x)} \cr 
          -\xi(x){\rm e}^{-i\theta(x)} & -\sigma\epsilon \cr}
  \pmatrix{\psi_4(x)\cr \psi_2(x)\cr}.   \label{eq:Dirac5-2}
\end{eqnarray}
This is the equation we shall numerically analyze.\par
Before investigating the Dirac equation, let us write down the
chiral currents in our basis. 
They are defined by
\begin{eqnarray*}
  j_L^\mu(x) &=& \bar\psi(x)\gamma^\mu{{1-\gamma_5}\over2}\psi(x), \\
  j_R^\mu(x) &=& \bar\psi(x)\gamma^\mu{{1+\gamma_5}\over2}\psi(x).
\end{eqnarray*}
It is easy to see that the relevant currents are written as
\begin{eqnarray}
  j_L^3(x) &=& \abs{\psi_4(x)}^2 - \abs{\psi_3(x)}^2, \label{eq:def-Lcurr}\\
  j_R^3(x) &=& \abs{\psi_1(x)}^2 - \abs{\psi_2(x)}^2. \label{eq:def-Rcurr}
\end{eqnarray}
The asymptotic forms of these currents will be used to calculate
the transmission and reflection coefficients.
\subsection{Extraction of the oscillating factor}
Here, for definiteness, we consider a scattering state in which
the incident wave is coming from the symmetric phase region 
($x\sim-\infty$) so that there is only the transmitted wave (right moving)
in the broken phase ($x\sim+\infty$).
This boundary condition is expressed as
$$
  \psi_a(x) \propto {\rm e}^{\sigma\alpha x}\qquad\mbox{as }
        x\rightarrow+\infty,
$$
where $\alpha = i\sqrt{\epsilon^2-\xi^2}$ and $a$ denote the spinor 
index.
Our aim is to find out the coefficients of the right and left movers
in the symmetric phase region of this state, from which
we can evaluate the chiral transmission and reflection 
coefficients.\par
For this purpose, we numerically integrate the Dirac equation starting 
from the plane wave in the deep broken phase, $x=+\infty$.
However it is rather difficult to implement this situation 
numerically, since preparing the plane wave at $x=\infty$ is impossible,
while it is the eigenstate only there.
One may think that an appropriate change of the variable which brings
$x=+\infty$ to some finite coordinate makes the wave function rapidly
oscillate near the boundary of the region.
To remedy this, we factorize the oscillating part of the wave function.
\par
Before performing the factorization, let us remove the effect of CP
violation in the broken phase by a unitary transformation of the
two-spinor. 
Introducing a unitary matrix
\begin{equation}
 U={1\over\sqrt{2}}\pmatrix{i&-i\cr
                           {\rm e}^{i\theta_0}&{\rm e}^{i\theta_0}\cr},
         \label{eq:def-U}
\end{equation}
and transforming the spinor as
\begin{equation}
 \pmatrix{\tpsi_1(x)\cr \tpsi_3(x)\cr} =
  U^{-1}\pmatrix{\psi_1(x)\cr \psi_3(x)\cr},\qquad
 \pmatrix{\tpsi_4(x)\cr \tpsi_2(x)\cr} =
  U^{*-1}\pmatrix{\psi_4(x)\cr \psi_2(x)\cr}, \label{eq:def-tpsi}
\end{equation}
the Dirac equation is reduced to
\begin{eqnarray}
 \del_x\pmatrix{\tpsi_1(x)\cr \tpsi_3(x)\cr} &=&
 \pmatrix{\xi f(x) & -i[\sigma\epsilon+\xi g(x)] \cr
          -i[\sigma\epsilon-\xi g(x)] & -\xi f(x) \cr}
 \pmatrix{\tpsi_1(x)\cr \tpsi_3(x)\cr},       \label{eq:Dirac6-1}\\
 \del_x\pmatrix{\tpsi_4(x)\cr \tpsi_2(x)\cr} &=&
 \pmatrix{-\xi f(x) & -i[\sigma\epsilon+\xi g(x)] \cr
          -i[\sigma\epsilon-\xi g(x)] & \xi f(x) \cr}
 \pmatrix{\tpsi_4(x)\cr \tpsi_2(x)\cr},          \label{eq:Dirac6-2}
\end{eqnarray}
where
\begin{equation}
 \xi f(x)\equiv\xi(x)\cos(\theta(x)-\theta_0), \qquad
 \xi g(x)\equiv\xi(x)\sin(\theta(x)-\theta_0).    \label{eq:def-fg}
\end{equation}
These satisfy the boundary conditions
\begin{eqnarray*}
 f(x)&\rightarrow&1,\quad g(x)\rightarrow0,\qquad
     \mbox{as }x\rightarrow+\infty\\
 f(x)&\rightarrow&0,\quad g(x)\rightarrow0,\qquad
     \mbox{as }x\rightarrow-\infty.
\end{eqnarray*}
\par
In the following, we consider only the positive frequency wave 
($\sigma=+$) for definiteness.
For the boundary condition we concern, the wave functions behave as,
at $x\sim+\infty$,
\begin{equation}
  \tpsi_a(x) \sim C_a{\rm e}^{\alpha x}.       \label{eq:asymp-tpsi}
\end{equation}
The asymptotic form of (\ref{eq:Dirac6-1}) yields
$$
 (\xi-\alpha)C_1{\rm e}^{\alpha x}= i\epsilon C_3{\rm e}^{\alpha x},
 \qquad
 (\xi+\alpha)C_3{\rm e}^{\alpha x}=-i\epsilon C_1{\rm e}^{\alpha x},
$$
which gives
\begin{equation}
  C_3 = {{\xi-\alpha}\over{i\epsilon}}C_1.      \label{eq:C1-3}
\end{equation}
Similarly we have
\begin{equation}
  C_4 = {{\xi-\alpha}\over{i\epsilon}}C_2.      \label{eq:C2-4}
\end{equation}
Now we factorize the wave function as
\begin{equation}
 \begin{array}{l}
  \displaystyle{\tpsi_1(x) = C_1{\rm e}^{\alpha x}\chi_1(x),}\\
  \displaystyle{\tpsi_2(x) = C_2{\rm e}^{\alpha x}\chi_2(x),}\\
  \displaystyle{\tpsi_3(x) ={{\xi-\alpha}\over{i\epsilon}}C_1
                             {\rm e}^{\alpha x}\chi_3(x),}\\
  \displaystyle{\tpsi_4(x) ={{\xi-\alpha}\over{i\epsilon}}C_2
                             {\rm e}^{\alpha x}\chi_4(x).}
 \end{array}  \label{eq:def-chi}
\end{equation}
The boundary condition is now
\begin{equation}
  \chi_a(x=+\infty) = 1.          \label{eq:bc-chi}
\end{equation}
In terms of $\chi_a$, the Dirac equation (\ref{eq:Dirac6-1}) and
(\ref{eq:Dirac6-2}) is expressed as
\begin{eqnarray}
 \begin{array}{l}
  \displaystyle{\chi_1^\prime(x) =
   (\xi-\alpha)(\chi_1(x)-\chi_3(x))+\xi(f(x)-1)\chi_1(x)
   -{{\xi(\xi-\alpha)}\over\epsilon}g(x)\chi_3(x),}   \\
  \displaystyle{\chi_3^\prime(x) =
   (\xi+\alpha)(\chi_1(x)-\chi_3(x))-\xi(f(x)-1)\chi_3(x)
   -{{\xi(\xi+\alpha)}\over\epsilon}g(x)\chi_1(x),}   \\
 \end{array}   \label{eq:Dirac7-1}\\
 \begin{array}{l}
  \displaystyle{\chi_2^\prime(x) =
   (\xi-\alpha)(\chi_2(x)-\chi_4(x))+\xi(f(x)-1)\chi_2(x)
   +{{\xi(\xi-\alpha)}\over\epsilon}g(x)\chi_4(x),}   \\
  \displaystyle{\chi_4^\prime(x) =
   (\xi+\alpha)(\chi_2(x)-\chi_4(x))-\xi(f(x)-1)\chi_4(x)
   +{{\xi(\xi+\alpha)}\over\epsilon}g(x)\chi_2(x).}
 \end{array}   \label{eq:Dirac7-2}
\end{eqnarray}
Note that the right-hand sides of the above equations vanish at
$x=+\infty$.
\subsection{Reflection and transmission coefficients}
To determine the transmission and reflection
coefficients, we need their asymptotic forms.
Suppose that the wave function behaves as, at $x\sim-\infty$,
\begin{equation}
  \psi_a(x)\sim A_a{\rm e}^{\beta x}+B_a{\rm e}^{-\beta x},
       \label{eq:asym-psi}
\end{equation}
where $\beta=i\epsilon$.
Then the incident and reflection currents are given by
\begin{equation}
 \begin{array}{l}
  \displaystyle{
   \left(j^3_L\right)^{\rm inc} = \abs{A_4}^2-\abs{A_3}^2,\qquad
   \left(j^3_R\right)^{\rm inc} = \abs{A_1}^2-\abs{A_2}^2,}\\
  \displaystyle{
   \left(j^3_L\right)^{\rm refl} = \abs{B_4}^2-\abs{B_3}^2,\qquad
   \left(j^3_R\right)^{\rm refl} = \abs{B_1}^2-\abs{B_2}^2.}
 \end{array}                    \label{eq:curr-AB}
\end{equation}
In the asymptotic region ($x\sim-\infty$), the coefficients
$A_a$ and $B_a$ are given by
\begin{eqnarray*}
 A_a&=&{{{\rm e}^{-\beta x}}\over2}
        \left(\psi_a(x)+{1\over\beta}\psi_a^\prime(x)\right), \\
 B_a&=&{{{\rm e}^{\beta x}}\over2}
        \left(\psi_a(x)-{1\over\beta}\psi_a^\prime(x)\right).
\end{eqnarray*}
Strictly speaking, these are not independent of $x$ at any finite $x$.
However, we shall use these expressions to calculate the transmission 
and reflection coefficients for so large $x$ that they stay almost 
constant.\par
Because of (\ref{eq:def-tpsi}),
\begin{eqnarray*}
 \pmatrix{\psi_1\cr\psi_3\cr}&=& U\pmatrix{\tpsi_1\cr\tpsi_3\cr}=
 \pmatrix{ {i\over{\sqrt2}}(\tpsi_1-\tpsi_3) \cr
           {{{\rm e}^{i\theta_0}}\over{\sqrt2}}(\tpsi_1+\tpsi_3) \cr},\\
 \pmatrix{\psi_4\cr\psi_2\cr}&=& U^*\pmatrix{\tpsi_4\cr\tpsi_2\cr}=
 \pmatrix{ {i\over{\sqrt2}}(\tpsi_2-\tpsi_4) \cr
           {{{\rm e}^{-i\theta_0}}\over{\sqrt2}}(\tpsi_2+\tpsi_4) \cr}.
\end{eqnarray*}
In terms of $\chi_a(x)$, the ``constants'' are expressed as
\begin{equation}
 \begin{array}{l}
  \displaystyle{
  \abs{A_1}^2 = {{\abs{C_1}^2}\over8}
  \abs{ (1+{\alpha\over\beta})\chi_1(x)+{1\over\beta}\chi_1^\prime(x) -
       {{\xi-\alpha}\over\beta}\left[(1+{\alpha\over\beta})\chi_3(x)
                               +{1\over\beta}\chi_3^\prime(x)\right]}^2,}\\
  \displaystyle{
  \abs{A_3}^2 = {{\abs{C_1}^2}\over8}
  \abs{ (1+{\alpha\over\beta})\chi_1(x)+{1\over\beta}\chi_1^\prime(x) +
       {{\xi-\alpha}\over\beta}\left[(1+{\alpha\over\beta})\chi_3(x)
                               +{1\over\beta}\chi_3^\prime(x)\right]}^2,}\\
  \displaystyle{
  \abs{A_2}^2 = {{\abs{C_2}^2}\over8}
  \abs{ (1+{\alpha\over\beta})\chi_2(x)+{1\over\beta}\chi_2^\prime(x) +
       {{\xi-\alpha}\over\beta}\left[(1+{\alpha\over\beta})\chi_4(x)
                               +{1\over\beta}\chi_4^\prime(x)\right]}^2,}\\
  \displaystyle{
  \abs{A_4}^2 = {{\abs{C_2}^2}\over8}
  \abs{ (1+{\alpha\over\beta})\chi_2(x)+{1\over\beta}\chi_2^\prime(x) -
       {{\xi-\alpha}\over\beta}\left[(1+{\alpha\over\beta})\chi_4(x)
                               +{1\over\beta}\chi_4^\prime(x)\right]}^2,}
 \end{array}     \label{eq:A-chi}
\end{equation}
and
\begin{equation}
 \begin{array}{l}
  \displaystyle{
  \abs{B_1}^2 = {{\abs{C_1}^2}\over8}
  \abs{ (1-{\alpha\over\beta})\chi_1(x)-{1\over\beta}\chi_1^\prime(x) -
       {{\xi-\alpha}\over\beta}\left[(1-{\alpha\over\beta})\chi_3(x)
                               -{1\over\beta}\chi_3^\prime(x)\right]}^2,}\\
  \displaystyle{
  \abs{B_3}^2 = {{\abs{C_1}^2}\over8}
  \abs{ (1-{\alpha\over\beta})\chi_1(x)-{1\over\beta}\chi_1^\prime(x) +
       {{\xi-\alpha}\over\beta}\left[(1-{\alpha\over\beta})\chi_3(x)
                               -{1\over\beta}\chi_3^\prime(x)\right]}^2,}\\
  \displaystyle{
  \abs{B_2}^2 = {{\abs{C_2}^2}\over8}
  \abs{ (1-{\alpha\over\beta})\chi_2(x)-{1\over\beta}\chi_2^\prime(x) +
       {{\xi-\alpha}\over\beta}\left[(1-{\alpha\over\beta})\chi_4(x)
                               -{1\over\beta}\chi_4^\prime(x)\right]}^2,}\\
  \displaystyle{
  \abs{B_4}^2 = {{\abs{C_2}^2}\over8}
  \abs{ (1-{\alpha\over\beta})\chi_2(x)-{1\over\beta}\chi_2^\prime(x) -
       {{\xi-\alpha}\over\beta}\left[(1-{\alpha\over\beta})\chi_4(x)
                               -{1\over\beta}\chi_4^\prime(x)\right]}^2,}
 \end{array}      \label{eq:B-chi}
\end{equation}
where ${}^\prime$ denotes the derivative with respect to $x$.
On the other hand, the transmitted currents are
\begin{equation}
 \begin{array}{l}
 \displaystyle{
  \left(j^3_L\right)^{\rm trans} =
  \half\abs{C_2}^2\abs{1-{{\xi-\alpha}\over\beta}}^2 -
  \half\abs{C_1}^2\abs{1+{{\xi-\alpha}\over\beta}}^2,}\\
 \displaystyle{
  \left(j^3_R\right)^{\rm trans} =
  \half\abs{C_1}^2\abs{1-{{\xi-\alpha}\over\beta}}^2 -
  \half\abs{C_2}^2\abs{1+{{\xi-\alpha}\over\beta}}^2.}
 \end{array}            \label{eq:curr-C}
\end{equation}
Equipped with these coefficients, we are ready to evaluate the 
transmission and reflection coefficients.\par
The transmission and reflection coefficients of the chiral fermions 
are defined by
\begin{eqnarray}
 T_{L\rightarrow L(R)}&=& 
   \left.{{\left(j^3_{L(R)}\right)^{\rm trans}}\over
          {\left(j^3_L\right)^{\rm inc}}}
           \right|_{\left(j^3_R\right)^{\rm inc}=0},  \label{eq:def-T1}\\
 T_{R\rightarrow L(R)}&=& 
   \left.{{\left(j^3_{L(R)}\right)^{\rm trans}}\over
          {\left(j^3_R\right)^{\rm inc}}}
           \right|_{\left(j^3_L\right)^{\rm inc}=0},  \label{eq:def-T2}\\
 R_{R(L)\rightarrow L(R)}&=& 
  -\left.{{\left(j^3_{L(R)}\right)^{\rm refl}}\over
          {\left(j^3_{R(L)}\right)^{\rm inc}}}
           \right|_{\left(j^3_{L(R)}\right)^{\rm inc}=0}.\label{eq:def-R}
\end{eqnarray}
The condition $\left(j^3_{L(R)}\right)^{\rm inc}=0$ is realized by
appropriately choosing the ratio of $C_1$ and $C_2$.
For example, to realize $\left(j^3_R\right)^{\rm inc}=0$, we take
\begin{equation}
 \begin{array}{l}
  \displaystyle{
   \abs{C_1}^2 =
   \abs{ (1+{\alpha\over\beta})\chi_2(x)+{1\over\beta}\chi_2^\prime(x)+
       {{\xi-\alpha}\over\beta}\left[(1+{\alpha\over\beta})\chi_4(x)
                            +{1\over\beta}\chi_4^\prime(x)\right]}^2,}\\
  \displaystyle{
   \abs{C_2}^2 =
   \abs{ (1+{\alpha\over\beta})\chi_1(x)+{1\over\beta}\chi_1^\prime(x)-
       {{\xi-\alpha}\over\beta}\left[(1+{\alpha\over\beta})\chi_3(x)
                            +{1\over\beta}\chi_3^\prime(x)\right]}^2,}
 \end{array}                                          \label{eq:C-chi}
\end{equation}
so that $\left(j^3_R\right)^{\rm inc}=\abs{A_1}^2-\abs{A_2}^2=0$.
Given these $\abs{C_i}^2$, (\ref{eq:def-T1})$\sim$(\ref{eq:def-R})
yield the chiral transmission and reflection coefficients
by use of (\ref{eq:curr-AB}) and (\ref{eq:curr-C}).\par
\section{Numerical Analysis}
To realize the boundary condition (\ref{eq:bc-chi}) at $x=\infty$
on a computer, we change the variable of infinite range to
another of half-infinite range. For definiteness, we take
the following new variable
\begin{equation}
 y = {\rm e}^{-az} = {\rm e}^{-x} \in(0,\infty).   \label{eq:def-y}
\end{equation}
This transforms ${d\over{dx}}\mapsto-y{d\over{dy}}$ so that
it brings the singularity at $y=0$ into the Dirac equation.
As we show below, the right-hand sides of (\ref{eq:Dirac7-1})
and (\ref{eq:Dirac7-2}) well behave to cancel such singularity.\par
The bubble wall connecting the broken and the symmetric phase
is expected to behave as
\begin{equation}
 \begin{array}{ll}
   \displaystyle{\rho(z)\sim 1 + {\rm e}^{-bz},}&
   \qquad\mbox{as } z\rightarrow+\infty \\
   \displaystyle{\rho(z)\sim {\rm e}^{b^\prime z},}&
   \qquad\mbox{as } z\rightarrow-\infty
 \end{array}   \label{eq:asymp-rho}
\end{equation}
where $b(>0)$ and $b^\prime(>0)$ will be order of the Higgs masses.
Hence if we choose $a$ to be smaller than the Higgs mass ($a<b$),
the potentials in the Dirac equation behave as, at $y\sim0$
($x\sim+\infty$),
$$
  f(y)\sim 1 + y^\mu,\qquad g(y)\sim y^\nu,
$$
with $\mu,\nu>1$.
Now put
$$
 f(y) = 1+y^\mu\sum_{n=0}^\infty f_n y^n,\qquad
 g(y) = y^\nu\sum_{n=0}^\infty g_n y^n. \qquad
 (f_0\not=0, g_0\not=0)
$$
To find the asymptotic behaviors of $\chi_1(y)$ and $\chi_3(y)$,
expand them as
\begin{equation}
 \begin{array}{ll}
  \displaystyle{
  \chi_1(y) = 1 + y^\rho\sum_{n=0}^\infty a_n y^n,}\quad&
    ({\bf R}\ni\rho>0, a_0\not=0) \\
  \displaystyle{
  \chi_3(y) = 1 + y^\sigma\sum_{n=0}^\infty b_n y^n.}\quad&
    ({\bf R}\ni\sigma>0, b_0\not=0)
 \end{array}     \label{eq:asym-chi}
\end{equation}
Then (\ref{eq:Dirac7-1}) in terms of $y$ yield
\begin{eqnarray}
 -\sum_{n=0}^\infty(\rho+n)a_n y^n&=&
  (\xi-\alpha)\left(\sum_{n=0}^\infty a_n y^n
                 -y^{\sigma-\rho}\sum_{n=0}^\infty b_n y^n \right) 
        +\xi\,y^{\mu-\rho}\sum_{n=0}^\infty f_n y^n
     ( 1 + y^\rho\sum_{n=0}^\infty a_n y^n )     \nonumber\\
 & &
  -{{\xi(\xi-\alpha)}\over\epsilon}y^{\nu-\rho}\sum_{n=0}^\infty g_n y^n
    (1 + y^\sigma\sum_{n=0}^\infty b_n y^n ),   \label{eq:asym-eq1}\\
 -\sum_{n=0}^\infty(\sigma+n)b_n y^n&=&
  (\xi+\alpha)\left(y^{\rho-\sigma}\sum_{n=0}^\infty a_n y^n
                 -\sum_{n=0}^\infty b_n y^n \right)
        -\xi\,y^{\mu-\sigma}\sum_{n=0}^\infty f_n y^n
    ( 1 + y^\sigma\sum_{n=0}^\infty b_n y^n )    \nonumber\\
 & &
  -{{\xi(\xi+\alpha)}\over\epsilon}y^{\nu-\sigma}\sum_{n=0}^\infty g_n y^n
    (1 + y^\rho\sum_{n=0}^\infty a_n y^n ).   \label{eq:asym-eq2}
\end{eqnarray}
When $\mu<\rho,\sigma$ or $\nu<\rho,\sigma$, the right-hand sides of
the both equations have singularities.\footnote{
One may think that in (\ref{eq:asym-eq1}) when $\mu=\nu<\rho$, they cancel
with each other if $\epsilon f_0 = (\xi-\alpha)g_0$.
But this is impossible for real $f_0$ and $g_0$.}
When $\mu>\rho,\sigma$ and $\nu>\rho,\sigma$, we must have 
$\rho=\sigma$. This is because, otherwise, the right-hand side of
either (\ref{eq:asym-eq1}) or (\ref{eq:asym-eq2}) has a singularity, 
$b_0 y^{\sigma-\rho}$ or $a_0 y^{\rho-\sigma}$, which cannot vanish
because of $a_0\not=0$ and $b_0\not=0$.
Then $O(y^0)$ terms give
\begin{eqnarray*}
 (\xi-\alpha+\rho)a_0 - (\xi-\alpha)b_0 &=& 0,\\
 (\xi+\alpha)a_0 - (\xi+\alpha-\rho)b_0 &=& 0.
\end{eqnarray*}
To have nontrivial $a_0$ and $b_0$,
$$
 (\xi-\alpha+\rho)(\xi+\alpha-\rho)-(\xi-\alpha)(\xi+\alpha)
 = \rho(2\alpha-\rho)=0,
$$
which has no real solution $\rho>0$.
Therefore to avoid any singularity in the right-hand sides, we are left with
two possibilities:
\begin{equation}
 \rho=\sigma=\mu\le\nu,\qquad\mbox{or}\qquad \rho=\sigma=\nu\le\mu. 
       \label{eq:condition-on-exp}
\end{equation}
In these cases, the lowest order equations determine $a_0$ and $b_0$
in terms of $f_0$ and/or $g_0$. Needless to say, the same argument
also applies to the remaining components of the Dirac equation.\par
Thus the Dirac equations in terms of $y$ are free from any
singularity, so that they can be numerically integrated by use
of a standard algorithm such as the Runge-Kutta method.
Now we summarize our procedure to calculate the chiral reflection
coefficients:
\begin{enumerate}
 \item  Prepare the potential $f(y)$ and $g(y)$. These would be
 determined by the model we choose through the equations of motion.
 \item  Integrate the Dirac equation
 \begin{eqnarray}
  \begin{array}{l}
   \displaystyle{
   -y{{d\chi_1(y)}\over{dy}} =
    (\xi-\alpha)(\chi_1(y)-\chi_3(y))+\xi(f(y)-1)\chi_1(y)
    -{{\xi(\xi-\alpha)}\over\epsilon}g(y)\chi_3(y),}   \\
   \displaystyle{
   -y{{d\chi_3(y)}\over{dy}} =
    (\xi+\alpha)(\chi_1(y)-\chi_3(y))-\xi(f(y)-1)\chi_3(y)
    -{{\xi(\xi+\alpha)}\over\epsilon}g(y)\chi_1(y),}
  \end{array}                      \label{eq:Dirac8-1}\\
  \begin{array}{l}
   \displaystyle{
   -y{{d\chi_2(y)}\over{dy}} =
    (\xi-\alpha)(\chi_2(y)-\chi_4(y))+\xi(f(y)-1)\chi_2(y)
    +{{\xi(\xi-\alpha)}\over\epsilon}g(y)\chi_4(y),}  \\
   \displaystyle{
   -y{{d\chi_4(y)}\over{dy}} =
    (\xi+\alpha)(\chi_2(y)-\chi_4(y))-\xi(f(y)-1)\chi_4(y)
    +{{\xi(\xi+\alpha)}\over\epsilon}g(y)\chi_2(y),}
  \end{array}                      \label{eq:Dirac8-2}
 \end{eqnarray}
 starting from the boundary condition $\chi_a(y=0)=1$ to large
 $y$ with a step $\Delta y$.
 Within each step, one can control the precision by choosing
 sufficiently small meshes or using quality control algorithm
 which automatically divides the mesh until the desired precision
 is reached.
 \item  At the end of each step, evaluate the reflection and 
 transmission coefficients.
 \item Proceed to larger $y$, until the relative error in the
 coefficients comes to be smaller than the prescribed precision.
\end{enumerate}
We shall see how this program works in the next section.
\section{Examples}
Here we take the following potentials to illustrate the calculation
of the chiral reflection coefficients. The chiral charge flux can be
obtained by the procedure outlined in \cite{FKOTTb} integrating
the reflection coefficients multiplied by the statistical distribution
functions of the fermions.
\begin{itemize}
 \item[(i)] $\displaystyle{
             \xi(x){\rm e}^{i\theta(x)} =
             \xi{{1+\tanh x}\over2}
             \exp\left[i\Delta\theta{{1-\tanh x}\over2}\right].}$\\
             This with $\Delta\theta=-\pi$ was taken in \cite{NKC}.
 \item[(ii)] $\displaystyle{
              \xi(x){\rm e}^{i\theta(x)} =
              \xi{{1+\tanh x}\over2}{\rm e}^{i\theta(x)},}$\\
             with $\theta(x)$ being the solution to the equations of 
             motion of the gauge-Higgs system, depicted in Fig.~2
             of \cite{FKOTTc}, satisfying $\theta_0=\theta(y=0)=1$ and
             $\theta(y=\infty)=0$.
 \item[(iii)] $\displaystyle{
              \xi(x){\rm e}^{i\theta(x)} =
              \xi{{1+\tanh x}\over2}{\rm e}^{i\theta(x)},}$\\
             with $\theta(x)$ being the solution to the equations of 
             motion of the gauge-Higgs system, depicted in Fig.~3
             of \cite{FKOTTc}, satisfying 
             $\theta(y=0)=\theta(y=\infty)=0$.
\end{itemize}
In the charge transport scenario, the generated baryon number density 
is given by \cite{NKC}
\begin{equation}
 {{n_B}\over s} \simeq
 {\cal N}{{100}\over{\pi^2 g_*}}\,\kappa\alpha_W^4\,
  {{F_Y\tau}\over{u T^2}},      \label{eq:baryon-density}
\end{equation}
where ${\cal N}$ is the model-dependent factor of $O(1)$, $g_*$ is
the radiation degrees of freedom at $T_C$ and is of $O(100)$,
$\kappa\alpha_W^4$ is the sphaleron transition rate in the symmetric
phase and is of $O(10^{-6})$ and $u$ is the wall velocity.
$\tau$ stands for the transport time, which is given by the average
time the reflected fermion spends in the plasma before caught up with
the bubble wall. It is expected to be of $O(10)\sim O(1000)$ times
the thermal correlation lengths, depending on the wall velocity,
if the forward scattering is enhanced.
In the original scenario, the weak hypercharge was taken as the chiral
charge injected into the symmetric phase.
In the following, we mean by the chiral charge flux the dimensionless
quantity normalized as
\begin{equation}
    {{F_Q}\over{u T^3(Q_L-Q_R)}}.        \label{eq:normalized-flux}
\end{equation}
\par\noindent
$\underline{\mbox{\bf case (i)}}$\\
In terms of the variable $y$, the potentials are written as
\begin{equation}
 f(y)={1\over{1+y^2}}\cos\left({{\Delta\theta\,y^2}\over{1+y^2}}\right),
  \qquad
 g(y)={1\over{1+y^2}}\sin\left({{\Delta\theta\,y^2}\over{1+y^2}}\right).
                 \label{eq:f-g-i}
\end{equation}
Now we take $\Delta\theta=-\pi$. To show how the program in the previous
section works, we calculated the reflection and transmission 
coefficients of the left-handed fermion for $\xi=1$ and $\epsilon=1.2$.
The result is listed in Table~\ref{tab1} for first several $y$.
\begin{table}[hbtp]
\begin{center}
\begin{tabular}{|c||c|c|c|}
 \hline
 $y$&$R_{L\rightarrow R}$&$T_{L\rightarrow L}+T_{L\rightarrow R}$&unitarity\\
 \hline
20	&$5.42508153\times10^{-1}$&$4.57493406\times10^{-1}$&1.00000156\\
40	&$5.42731549\times10^{-1}$&$4.57268549\times10^{-1}$&1.00000010\\
60	&$5.42819216\times10^{-1}$&$4.57180804\times10^{-1}$&1.00000002\\
80	&$5.42830565\times10^{-1}$&$4.57169442\times10^{-1}$&1.00000001\\
100	&$5.42826265\times10^{-1}$&$4.57173738\times10^{-1}$&1.00000000\\
120	&$5.42820329\times10^{-1}$&$4.57179672\times10^{-1}$&1.00000000\\
140	&$5.42815576\times10^{-1}$&$4.57184425\times10^{-1}$&1.00000000\\
160	&$5.42812238\times10^{-1}$&$4.57187764\times10^{-1}$&1.00000000\\
180	&$5.42810021\times10^{-1}$&$4.57189980\times10^{-1}$&1.00000000\\
200	&$5.42808603\times10^{-1}$&$4.57191398\times10^{-1}$&1.00000000\\
220	&$5.42807730\times10^{-1}$&$4.57192270\times10^{-1}$&1.00000000\\
240	&$5.42807222\times10^{-1}$&$4.57192778\times10^{-1}$&1.00000000\\
260	&$5.42806955\times10^{-1}$&$4.57193046\times10^{-1}$&1.00000000\\
280	&$5.42806844\times10^{-1}$&$4.57193157\times10^{-1}$&1.00000000\\
300	&$5.42806833\times10^{-1}$&$4.57193167\times10^{-1}$&1.00000000\\
 \hline
\end{tabular}
\end{center}
\caption{The reflection and transmission coefficients of the 
left-handed fermion. These should become constants for $y\sim\infty$.}
\label{tab1}
\end{table}
The column `unitarity' means the sum of the two columns left to it. 
This shows that to get the precision of four digits, it is sufficient 
to integrate the equations up to $y=100$, which is amount to
$x = -\ln100\simeq -4.6052$ and is regarded far from the bubble wall.
In this case, one can obtain the same data of
$\Delta R = R_{R\rightarrow L}-R_{L\rightarrow R}$ as a function of
$(\epsilon,\xi)$ as shown in Fig.~2 of \cite{NKC}.\par
For $\Delta\theta=0.001$, we confirmed that our numerical method yields
the same $\Delta R$ as that obtained by the perturbative method in
\cite{FKOTTb}. For $\Delta\theta=-1$, which is equivalent to the profile
\begin{equation}
  \xi(x){\rm e}^{i\theta(x)} =
             \xi{{1+\tanh x}\over2}
             \exp\left[i{{1+\tanh x}\over2}\right],  \label{eq:ex-i}
\end{equation}
up to the unitary transformation (\ref{eq:def-tpsi}), we calculated
$\Delta R$ by the numerical method developed here. 
The chiral charge fluxes for various
mass and wall width are depicted in Fig.~1.\par\noindent
$\underline{\mbox{\bf case (ii)}}$\\
In this case, $\theta_0=1$ and the maximum of $g(x)$ is about $0.4$,
which is comparable with $f(x)$ at the same $x$.
So we did not evaluated the chiral charge flux in \cite{FKOTTc},
using our perturbative method\cite{FKOTTb}.
This profile, which is a solution to the equations of motion to
the gauge-Higgs system with the kink-type $\rho(x)$, indicates that
$\theta(x)$ is no longer proportional to the kink when $\theta_0$ is
$O(1)$. This should be compared with the profile (\ref{eq:ex-i}),
which satisfies the same boundary conditions.
We performed the numerical calculation of the chiral charge
flux, which is given in Fig.~2.
This shows that this profile slightly enhances the flux over the 
profile employed in \cite{NKC} by about a factor of 2.
The former is about $80\%$ larger near the peak,
and decreases more slowly for large mass or thick wall than the latter.
\par\noindent
$\underline{\mbox{\bf case (iii)}}$\\
Although the maximum of $\theta(x)$ reaches $0.37$, the potential
$f(x)$ has a kink-like shape and $g(x)$ is sufficiently small as
shown in Fig.~4 of \cite{FKOTTc}. 
We expected that the perturbative method is applicable to this case
and evaluated the chiral charge flux, depicted in Fig.~5 of \cite{FKOTTc}.
To confirm this, we calculated $\Delta R$ for several $(\epsilon, \xi)$.
One of the results is given in Fig.~3 for $\xi=0.4$ and $m_0=1$.
This shows that the perturbative method was a good approximation
in this case.
\section{Discussions}
We proposed a numerical method to solve the scattering problem of
the Dirac equation in the background of any CP-violating bubble wall.
By use of it, one can evaluate the reflection coefficients of
chiral fermions to any precision. This will be a powerful tool to
analyze electroweak baryogenesis based on the charge transport 
scenario.\par
Whether the charge transport or the spontaneous mechanism is effective
will be determined by the dynamics of the EWPT.
Recent study of the standard model at finite temperature\cite{EWPT}
suggests that the first order nature of the EWPT is stronger than
that expected from the one-loop calculation of the effective potential.
Even though efficient electroweak baryogenesis requires some extension
of the Higgs sector to incorporate spatially varying CP phase,
this fact would mean that the bubble wall would be thinner than
naively expected from the lowest order effective potential, so that
the charge transport scenario would play a crucial role in 
baryogenesis.\par
The profiles of the bubble wall which we took in this paper have the
kink-type $\rho(x)$.
In fact, they should be determined by the dynamics, and we have
shown that the modulus of the Higgs could be distorted from the kink,
when $\theta(x)$ is not small for the choice of the effective potential
which has a kink-type $\rho(x)$ for $\theta=0$\cite{FKOT}.
For such a profile, we can evaluate the flux precisely, by taking
sufficiently small $a$.\par
\vspace{40pt}
{\large\bf Figure Captions}\par\noindent
\begin{itemize}
\item[Fig.1:]  Contour plot of the chiral charge flux,
 $\log_{10}\left[-F_Q/(uT^3(Q_L-Q_R))\right]$ for the example (i) with
 $\theta_0=-1$.
 Here we take $u=0.1$ and $T=100$GeV. 
\item[Fig.2:] Contour plot of the chiral charge flux, normalized as
 $\log_{10}\left[-F_Q/(uT^3(Q_L-Q_R))\right]$ for the example (ii), 
 given in Fig.~2 of \cite{FKOTTc}.
 Here we take the same $u$ and $T$ as Fig.~1.
\item[Fig.3:] $\Delta R$ for the profile given in Fig.~3 of \cite{FKOTTc}.
 Here we take $\xi=0.4$ and $m_0=1$.
 The solid line represents the numerical result obtained here.
 The dashed line stands for the previous result based on the 
 perturbative method.
\end{itemize}
%
%
%
\epsfysize=\textheight
\centerline{\epsfbox{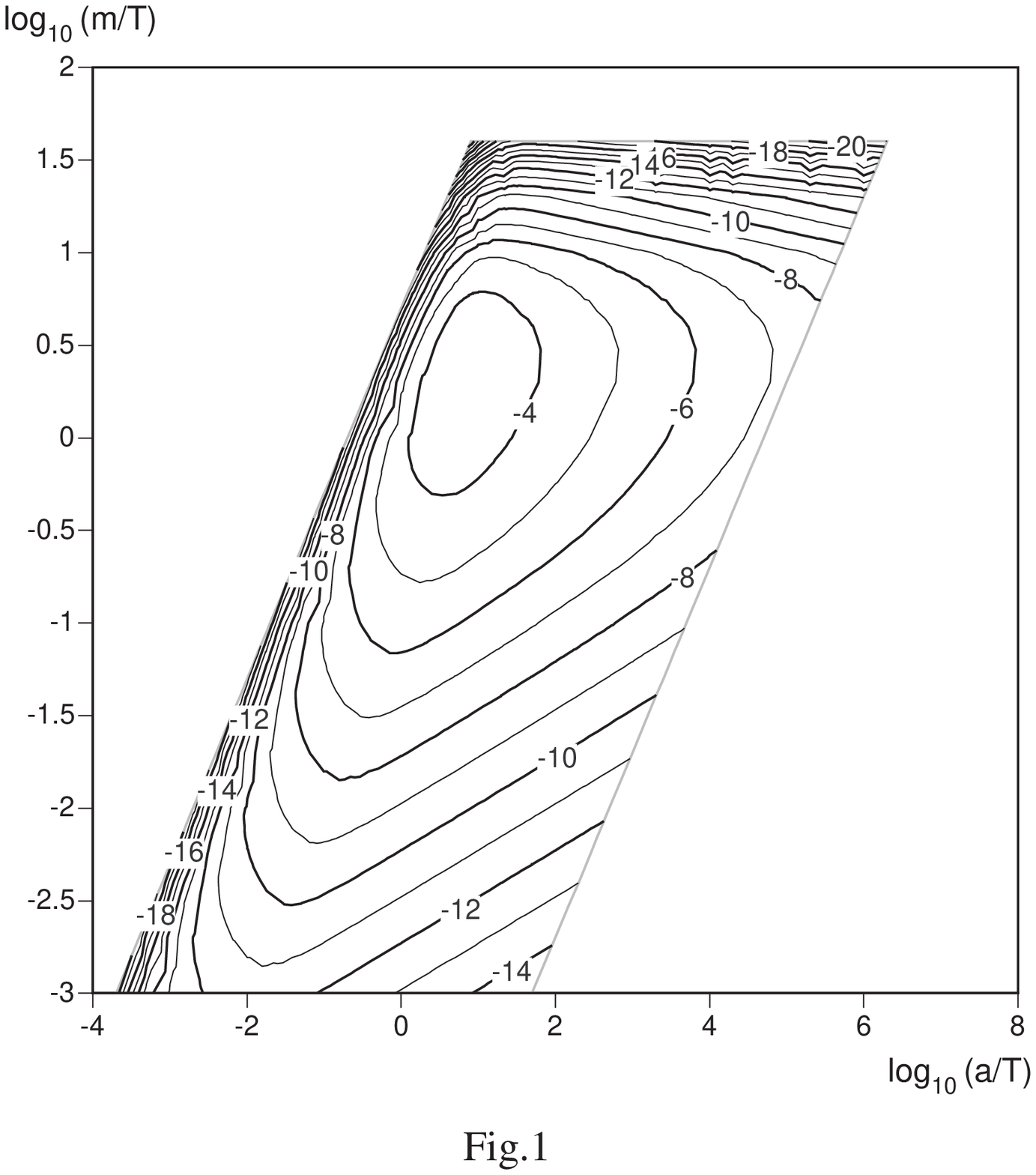}}
\newpage
\epsfysize=\textheight
\centerline{\epsfbox{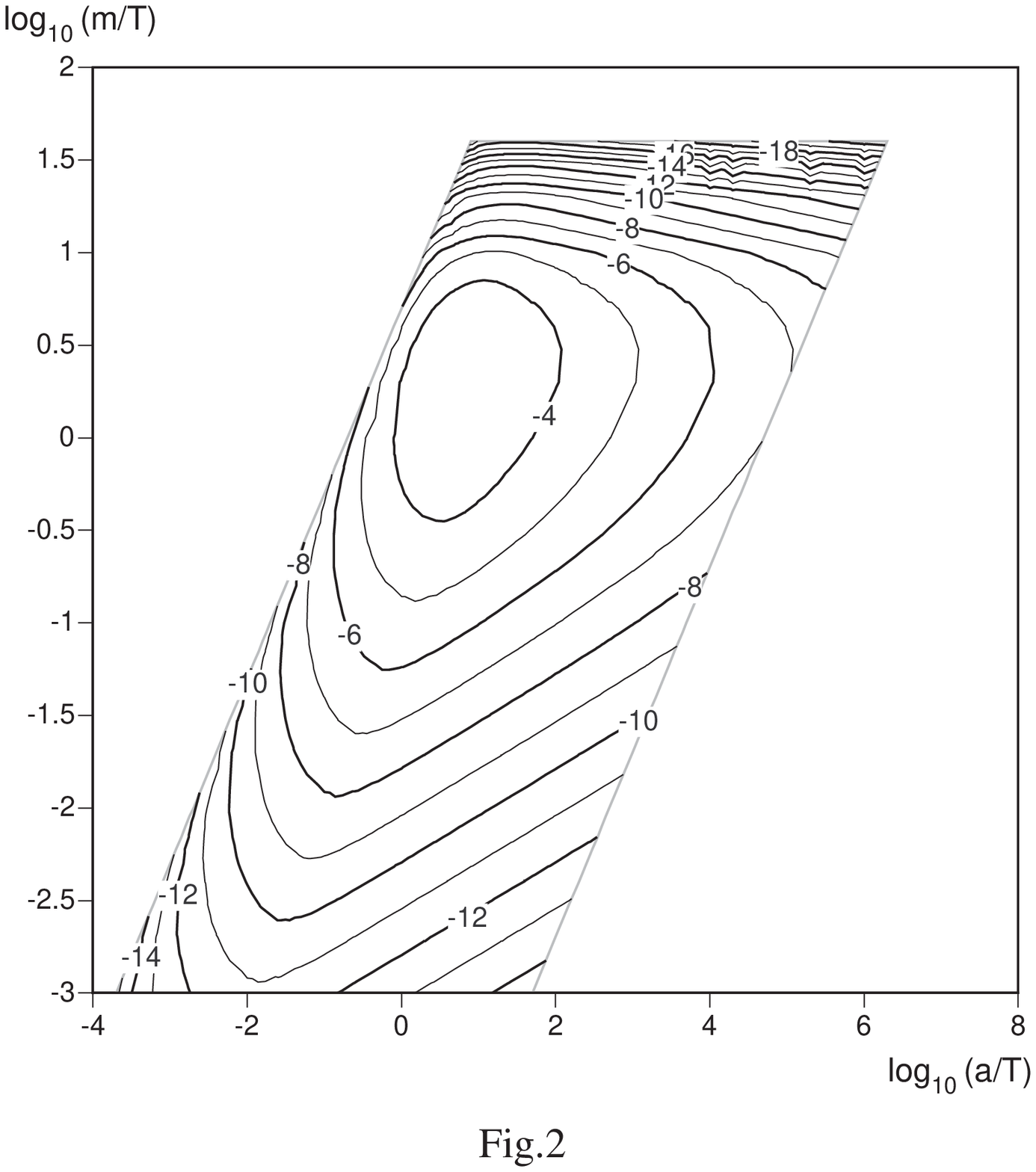}}
\newpage
\epsfysize=\textheight
\centerline{\epsfbox{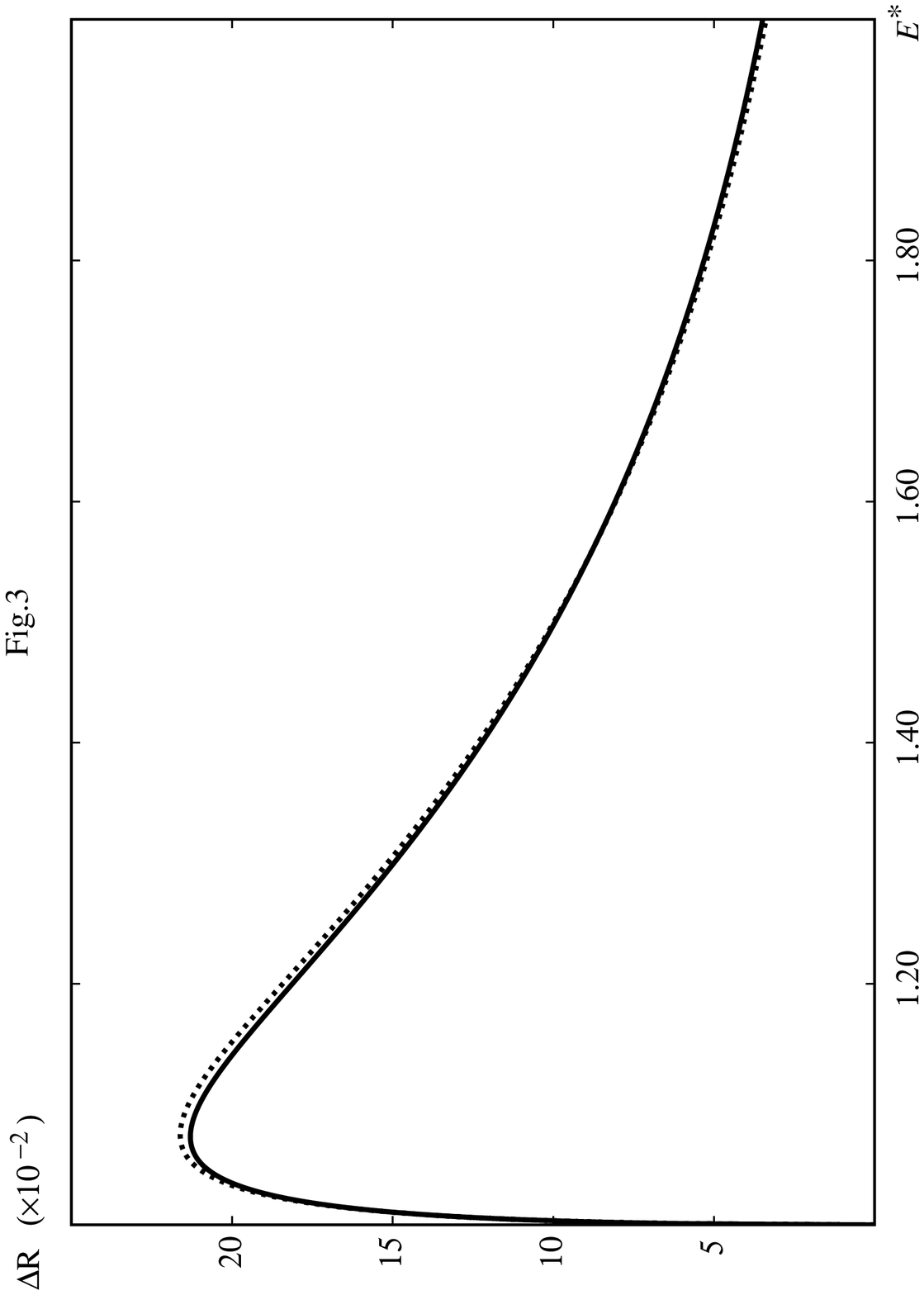}}
\end{document}